\documentclass[epj-spec]{svjour}
\usepackage{graphics}
\newcommand{\slim}{\mskip 1.5mu}              
\newcommand{\lf}{\left}
\newcommand{\rg}{\right}
\begin{document}
\title{New target transverse spin dependent azimuthal asymmetries from COMPASS experiment}
\author{B.~Parsamyan\inst{1,}\inst{2,}\thanks{\email{bakur.parsamyan@cern.ch}} (on behalf of the COMPASS collaboration)}

\institute{Dipartimento di Fisica Generale, Universit\`a di Torino
\and INFN, Sezione di Torino, Via P. Giuria 1, I-10125 Torino,
Italy}
\abstract{
In general, eight target transverse spin-dependent
azimuthal modulations are allowed in semi inclusive deep inelastic
scattering of polarized leptons on a transversely polarized target.
In the QCD parton model four of these asymmetries can be interpreted
within the leading order approach. Two of them, namely Collins and
Sivers effects were already measured by HERMES and COMPASS
experiments. Other two leading twist and remaining four azimuthal
asymmetries which can be interpreted as twist-three contributions
have been measured for the first time in COMPASS using a 160 GeV/c
longitudinally polarized ($P_{l}\simeq -0.8$) muon beam and a
transversely polarized $^6LiD$ target. We present here the
preliminary results from the 2002-2004 data.
} 
\maketitle
\section{Introduction}
\label{intro} Transverse spin effects in Semi Inclusive Deep
Inelastic Scattering (SIDIS) of polarized leptons on a transversely
polarized target have become an interesting issue in the past years.

The first two of the eight target transverse spin asymmetries which
are allowed in the general expression of the SIDIS cross-section
~\cite{Kotzinian:1994dv} and have been measured by HERMES and
COMPASS experiments ~\cite{herm,comp1,comp2} were the Collins and
the Sivers effects. Measurements done by these collaborations
together with the BELLE ~\cite{belle} data allowed for example, a
first extraction of the transversity and Sivers Transverse Momentum
Dependent (TMD) distribution functions (DFs) and Collins
fragmentation function (FF)
~\cite{Anselmino:2005ea},\cite{Anselmino:2007fs}. Here we present
preliminary results on the six remaining transverse spin asymmetries
first extracted by COMPASS from the 2002-2004 deuteron data.

\section{Definition of the asymmetries}\label{tr_spin_asym}

Based on the general principles of quantum field theory it can be
shown in a model independent way  that in the one photon exchange
approximation the cross-section of lepton-hadron SIDIS processes
include 18 structure functions ~\cite{Kotzinian:1994dv},
\cite{Bacchetta:2006tn}:
\begin{eqnarray}\label{e:crossmaster}
&&\frac{d\sigma}{dx \, dy\, d\psi \,dz\, d\phi_h\, d
P_{hT}^2} = \frac{\alpha^2}{x y\slim
Q^2}\, \frac{y^2}{2\,(1-\varepsilon)}\, \biggl( 1+\frac{\gamma^2}{2x}
\biggr)\, \times \nonumber \\
&& \quad \times\, \Biggl\{ F_{UU ,T} + \varepsilon\slim F_{UU ,L} +
\sqrt{2\,\varepsilon (1+\varepsilon)}\,\cos\phi_h\,
F_{UU}^{\cos\phi_h} \nonumber \\
&& \quad +\, \varepsilon
\cos(2\phi_h)\, F_{UU}^{\cos 2\phi_h} + P_{l}\,
\sqrt{2\,\varepsilon (1-\varepsilon)}\,
           \sin\phi_h\,
F_{LU}^{\sin\phi_h} \phantom{\Bigg[ \Bigg] } \nonumber \\
&& \quad +\, S_L\, \Bigg[
 \sqrt{2\, \varepsilon (1+\varepsilon)}\,
  \sin\phi_h\,
F_{UL}^{\sin\phi_h} +  \varepsilon \sin(2\phi_h)\, F_{UL}^{\sin
2\phi_h} \Bigg] \nonumber \\
&&\quad +\, S_L P_{l}\, \Bigg[
\,\sqrt{1-\varepsilon^2}\;F_{LL} +\sqrt{2\,\varepsilon (1-\varepsilon)}\,
\cos\phi_h\,F_{LL}^{\cos \phi_h} \Bigg] \nonumber \\
&& \quad +\,|{\bf S}_T|\, \Bigg[ \sin(\phi_h-\phi_S)\, \Bigl(F_{UT
,T}^{\sin\lf(\phi_h -\phi_S\rg)} + \varepsilon\,
F_{UT,L}^{\sin\lf(\phi_h -\phi_S\rg)}\Bigr)\\
&& \quad +\, \varepsilon\, \sin(\phi_h+\phi_S)\,
F_{UT}^{\sin\lf(\phi_h +\phi_S\rg)} + \varepsilon\,
\sin(3\phi_h-\phi_S)\, F_{UT}^{\sin\lf(3\phi_h -\phi_S\rg)}
\phantom{\Bigg[ \Bigg] }\nonumber\\
&& \quad +\, \sqrt{2\,\varepsilon (1+\varepsilon)}\, \sin\phi_S\,
F_{UT}^{\sin \phi_S } + \sqrt{2\,\varepsilon (1+\varepsilon)}\,
\sin(2\phi_h-\phi_S)\, F_{UT}^{\sin\lf(2\phi_h -\phi_S\rg)} \Bigg]
\nonumber\\
&& \quad +\, |{\bf S}_T| P_{l}\, \Bigg[\sqrt{1-\varepsilon^2}\,
\cos(\phi_h-\phi_S)\, F_{LT}^{\cos(\phi_h -\phi_S)}\nonumber \\
&& \quad +\, \sqrt{2\,\varepsilon
(1-\varepsilon)}\, \cos\phi_S\,F_{LT}^{\cos \phi_S} + \sqrt{2\,\varepsilon (1-\varepsilon)}\,\cos(2\phi_h-\phi_S)\,
F_{LT}^{\cos(2\phi_h - \phi_S)} \Bigg] \Biggr\}\nonumber,
\end{eqnarray}
where the standard SIDIS notations are used, and the ratio
$\varepsilon$ of longitudinal and transverse photon fluxes is given
by
\begin{equation} \label{eq:eps}
\varepsilon = \frac{1-y -\frac{1}{4}\slim \gamma^2 y^2}{1-y
  +\frac{1}{2}\slim y^2 +\frac{1}{4}\slim \gamma^2 y^2},
\end{equation}
where $\gamma = \frac{2 M x}{Q}$. The notations for the structure
functions $F_{sub}^{sup}$ which on the r.h.s.\ depend on $x$, $Q^2$,
$z$ and $P_{hT}$ have the following meaning: the superscript
corresponds to the azimuthal asymmetry described by the given
structure function, whereas the first and second subscripts indicate
the respective ("U"-unpolarized,"L"-longitudinal and "T"-transverse)
polarization of beam and target and the third one specifies the
polarization of the virtual photon. Integrating these structure
functions over the produced hadron momentum and summing over all
hadrons in the final state one can find relations between the
polarized SIDIS structure functions and ordinary DIS structure
functions. For more details see ~\cite{Kotzinian:1994dv},
\cite{Bacchetta:2006tn}.

Azimuthal angles have the following notations: $\phi_h$ is the
azimuthal angle of the produced hadron, $\phi_S$ of the nucleon spin
and $\psi$ is the laboratory azimuthal angle of the scattered
lepton, and in DIS kinematics $d\psi \approx d\phi_S$.

As one can see from expression Eq.~(\ref{e:crossmaster}), there are
only eight target transverse polarization dependent azimuthal
modulations:
\begin{eqnarray}
&&w_1(\phi_h, \phi_s)=\sin(\phi _h -\phi _s ),\nonumber \\
&&w_2(\phi_h,\phi_s)=\sin(\phi _h +\phi _s ),\nonumber \\
&&w_3(\phi_h,\phi_s)=\sin(3\phi _h -\phi _s ),\nonumber \\
&&w_4(\phi_h,\phi_s)=\sin(\phi _s ),\\
&&w_5(\phi_h,\phi_s)=\sin(2\phi_h -\phi _s ),\nonumber \\
&&w_6(\phi_h,\phi_s)=\cos(\phi _h -\phi _s ),\nonumber \\
&&w_7(\phi_h,\phi_s)=\cos(\phi _s ),\nonumber \\
&&w_8(\phi_h,\phi_s)=\cos(2\phi_h -\phi _s )\nonumber
\end{eqnarray}
Five of them are single target spin dependent and three are double
beam-target spin dependent asymmetries. The first two modulations
$w_1(\phi_h, \phi_s)$ and $w_2(\phi_h, \phi_s)$ correspond to the
Sivers and Collins effects.

The expression for the cross section can be represented in terms of
the asymmetries:
\begin{eqnarray}\label{eq:cros-sect-short}
    d \sigma(\phi_h,\phi_s,...) &\propto& (1 +
    |{\bf S}_T| {\sum_{i=1}^5} D^{w_i(\phi_h, \phi_s)} A_{UT}^{w_i(\phi_h,
\phi_s)}
    w_i(\phi_h,\phi_s)\\ \nonumber
    & + &
    P_{l} |{\bf S}_T| {\sum_{i=6}^8} D^{w_i(\phi_h, \phi_s)}
    A_{LT}^{w_i(\phi_h, \phi_s)} w_i(\phi_h,\phi_s) + ...\big),
\end{eqnarray}
where ${\bf S}_T$ is the target transverse polarization and $P_l$ is the beam polarization. The depolarization factors $D^{w_i(\phi_h, \phi_s)}$, have been factored out, and the asymmetries have been defined as the ratios of
corresponding structure functions to the unpolarized one:

\begin{equation}\label{eq:as_def}
    A_{BT}^{w_i(\phi_h, \phi_s)} \equiv  \frac{F_{BT}^{w_i(\phi_h,
    \phi_s)}}{F_{UU,T}},
\end{equation}
where $B=L$ or $B=U$ indicates the beam polarization.

The depolarization factors entering in Eq.~(\ref{eq:cros-sect-short}) depend only on $y$ and are given as
\begin{eqnarray}\label{eq:depol}
    && D^{\sin(\phi _h -\phi _s )}(y) = 1,\;\;
    D^{\cos(\phi _h -\phi _s )}(y) =\frac {y(2-y)} {1+(1-y)^2},\nonumber\\
    && D^{\sin(\phi _h +\phi _s )}(y) = D^{\sin(3\phi _h +\phi _s )}(y)
=\frac {2(1-y)} {1+(1-y)^2}, \\
    && D^{\sin(2\phi _h -\phi _s )}(y) = D^{\sin(\phi _s )}(y) = \frac
{2(2-y)\sqrt{1-y}} {1+(1-y)^2}, \nonumber \\
    &&D^{\cos(2\phi _h -\phi _s )}(y) = D^{\cos(\phi _s )}(y) = \frac
{2y\sqrt{1-y}}
    {1+(1-y)^2}.\nonumber
\end{eqnarray}

The asymmetries extracted from the data as the amplitudes of the
corresponding azimuthal modulations (raw asymmetries) are then given
by
\begin{eqnarray}\label{eq:as_exp}
A_{UT, \; raw}^{w_i(\phi_h, \phi_s)} &=&D^{w_i(\phi_h, \phi_s)}(y) f
|S_T| A_{UT}^{w_i(\phi_h, \phi_s)}\, \;(i=1,5),  \\
A_{LT, \; raw}^{w_i(\phi_h, \phi_s)} &=&D^{w_i(\phi_h, \phi_s)}(y) f
P_{l} |S_T| A_{LT}^{w_i(\phi_h, \phi_s)}\, \;(i=6,8)
\end{eqnarray}
where $f$ is the target polarization dilution factor.

In the QCD parton model four of the eight transverse asymmetries are
given by the ratio of convolutions of spin-dependent to
spin-independent twist two DFs and FFs:

\begin{eqnarray}
&&A_{UT}^{\sin (\phi _h -\phi _s )} \propto \frac{f_{1T}^{\bot q} \otimes
D_{1q}^h}{f_1^q \otimes D_{1q}^{h}}, \nonumber \\
&&A_{UT}^{\sin (\phi _h +\phi _s )} \propto \frac{h_1^q \otimes H_{1q}^{\bot
h}}{f_1^q \otimes D_{1q}^{h}}, \\
&&A_{LT}^{\cos (\phi _h -\phi _s )} \propto \frac{g_{1T}^q \otimes
D_{1q}^h}{f_1^q \otimes D_{1q}^{h}}, \nonumber \\
&&A_{UT}^{\sin (3\phi _h
-\phi _s )} \propto \frac{h_{1T}^{\bot q} \otimes H_{1q}^{\bot
h}}{f_1^q \otimes D_{1q}^{h}} \nonumber
\label{eq:LO_as}
\end{eqnarray}

As an example, the $A_{LT}^{\cos (\phi _h -\phi _s )}$ and
$A_{UT}^{\sin (3\phi _h -\phi _s )}$ leading-twist asymmetries can
be used for extraction of DFs $g_{1T}^q$ and $h_{1T}^{\perp\,q}$
describing the quark longitudinal and transverse (along the quark
transverse momentum) polarization in the transversely polarized
nucleon. The other four asymmetries can be interpreted as Cahn
kinematic corrections to spin effects on the transversely polarized
nucleon ~\cite{Kotzinian:1994dv}:

\begin{eqnarray}
&&A_{LT}^{\cos (\phi _s )} \propto \frac{M}{Q}\frac{g_{1T}^q \otimes
D_{1q}^h}{f_1^q \otimes D_{1q}^{h}}, \nonumber \\
&&A_{LT}^{\cos (2\phi _h -\phi _s )} \propto
\frac{M}{Q}\frac{g_{1T}^q \otimes D_{1q}^h}{f_1^q \otimes D_{1q}^{h}},\\
&&A_{UT}^{\sin (\phi _s )} \propto \frac{M}{Q}\frac{{h_1^q \otimes
H_{1q}^{\bot h} +f_{1T}^{\bot q} \otimes D_{1q}^h }}{f_1^q \otimes D_{1q}^{h}}, \nonumber\\
&&A_{UT}^{\sin (2\phi _h -\phi _s )} \propto
\frac{M}{Q}\frac{{h_{1T}^{\bot q} \otimes H_{1q}^{\bot h}
+f_{1T}^{\bot q} \otimes D_{1q}^h }}{f_1^q \otimes D_{1q}^{h}}.\nonumber
 \label{eq:subLO_as}
\end{eqnarray}

\section{Analysis method and results}\label{Analysis}

In this section we will briefly review the analysis method used in
COMPASS for the extraction of transverse spin asymmetries. The event
selection procedure and the analysis method are the same as the one
applied for already published Collins and Sivers asymmetries, and a
detailed description can be found in ~\cite{comp2}.

In our analysis we used the COMPASS data collected in years
2002-2004 with the 160 GeV/c longitudinally polarized
($P_{l}\simeq -0.8$) muon beam and a transversely polarized
$^6LiD$ target. The COMPASS target consists of two oppositely
polarized target cells with the dilution factor $\simeq 0.38$ and
average polarization $\simeq 50\%$. Once per week the polarization
was reversed in both cells. Such a configuration of the target
serves to reduce the systematic effects arising due to the
difference in acceptance of the target cells.

The following kinematic cuts were imposed in the analysis: $Q^2>1$
(GeV/c)$^2$, $W>5$ GeV, $0.1<y<0.9$, $P_T^h>0.1$ GeV/c and $z>0.2$.

 One can see that eight transverse spin
modulations are based on five combinations of azimuthal hadron ($
\phi_h$) and spin ($ \phi_s$) angles which are: $ \Phi_1 = \phi_h -
\phi_s, \, \Phi_2 = \phi_h + \phi_s, \, \Phi_3 = 3\phi_h - \phi_s,
\, \Phi_4 = \phi_s, \, \Phi_5 = 2\phi_h - \phi_s$.Therefore, we can
define the following five $ \Phi_j $ dependent modulations:
\begin{eqnarray}\label{eq:W_desc}
    && W_1 (\Phi_1)  = A^{w_1(\phi_h, \phi_s)}_{raw}  \sin(\Phi_1) + A^{w_6(\phi_h, \phi_s)}_{raw}  \cos(\Phi_1) \nonumber\\
    && W_2 (\Phi_2)  = A^{w_2(\phi_h, \phi_s)}_{raw}  \sin(\Phi_2) \nonumber \\
    && W_3 (\Phi_3)  = A^{w_3(\phi_h, \phi_s)}_{raw}  \sin(\Phi_3) \\
    && W_4 (\Phi_4)  = A^{w_4(\phi_h, \phi_s)}_{raw}  \sin(\Phi_4) + A^{w_7(\phi_h, \phi_s)}_{raw}  \cos(\Phi_4)\nonumber\\
    && W_5 (\Phi_5)  = A^{w_5(\phi_h, \phi_s)}_{raw}  \sin(\Phi_5) + A^{w_8(\phi_h, \phi_s)}_{raw}  \cos(\Phi_5)\nonumber
\end{eqnarray}

For each subperiod of our measurement and each target cell, we can
now describe the $ \Phi_j $ distribution of the number of events by
\begin{equation}\label{eq:cs_desc}
N^{\pm}_{u/d}(\Phi_j) = F^{\pm}_{u/d} n^{\pm}_{u/d}
a^{\pm}_{u/d}(\Phi_j) \sigma ( 1 \pm W_j(\Phi_j))
\end{equation}
where +(-) indicates up (down) target polarization and u(d) the
upstream and downstream target cells, $ \sigma$ is the unpolarized
cross-section, $F^{\pm}_{u/d}$ is the flux and $n^{\pm}_{u/d}$ the
target density. Finally, $  a^{\pm}_{u/d}(\Phi_j) $ is the $ \Phi_j
$ dependent acceptance for the corresponding cell and polarization
state.

\begin{figure}[ht!]
\center \resizebox{0.85\textwidth}{!} {
\includegraphics{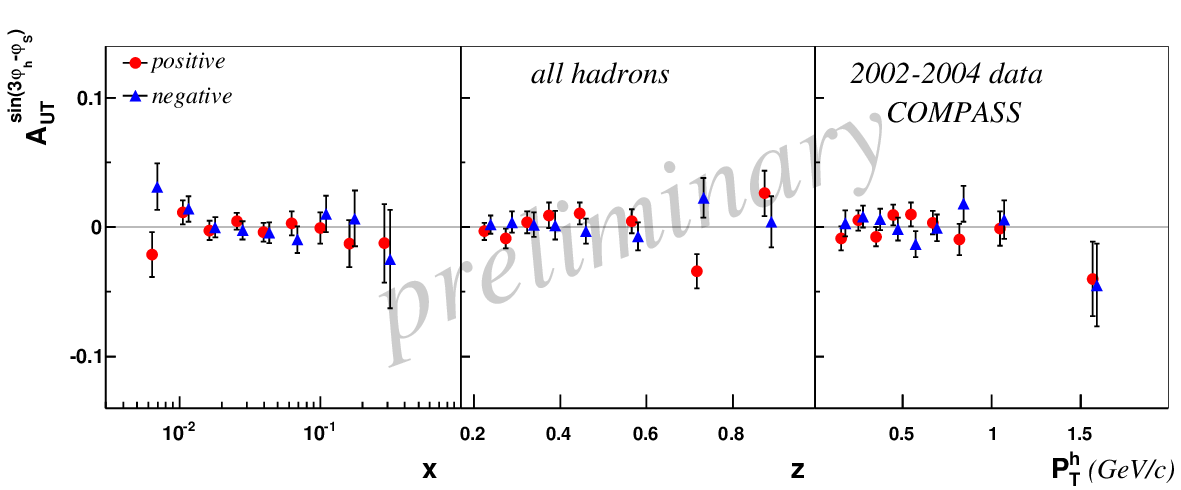} }
\caption{$A_{UT}^{sin(3\phi_h-\phi_s)}$ asymmetry for positive (red
circles) and negative (blue triangles) hadrons vs. $\it x$, $z$ and
$P_{hT}$.}
\label{fig:1}       
\end{figure}

\begin{figure}[ht!]
\center \resizebox{0.85\textwidth}{!} {
\includegraphics{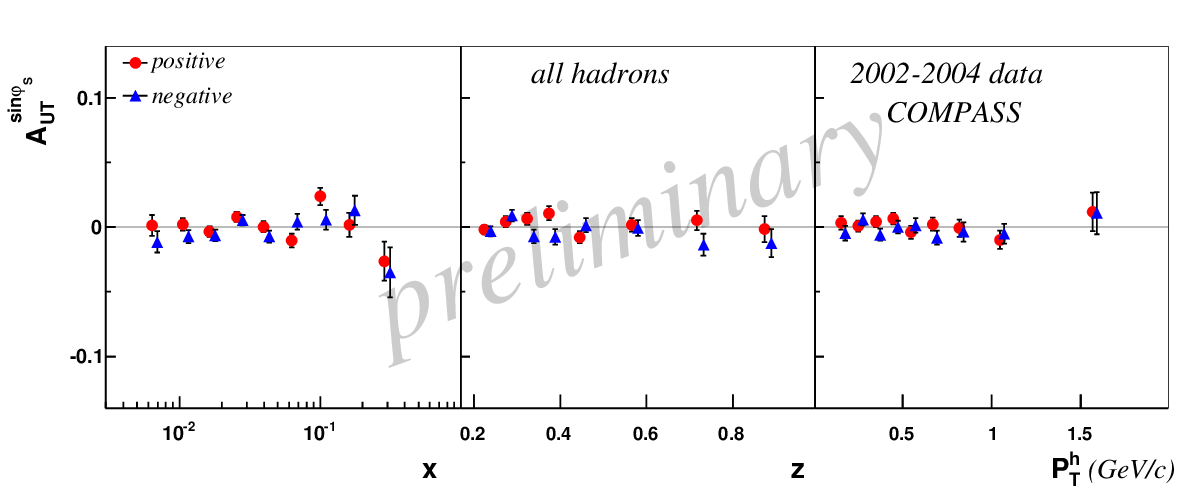} } \caption{$A_{UT}^{sin\phi_s}$
asymmetry for positive (red circles) and negative (blue triangles)
hadrons vs. $\it x$, $z$ and $P_{hT}$.}
\label{fig:2}       
\end{figure}

\begin{figure}[ht!]
\center \resizebox{0.85\textwidth}{!} {
\includegraphics{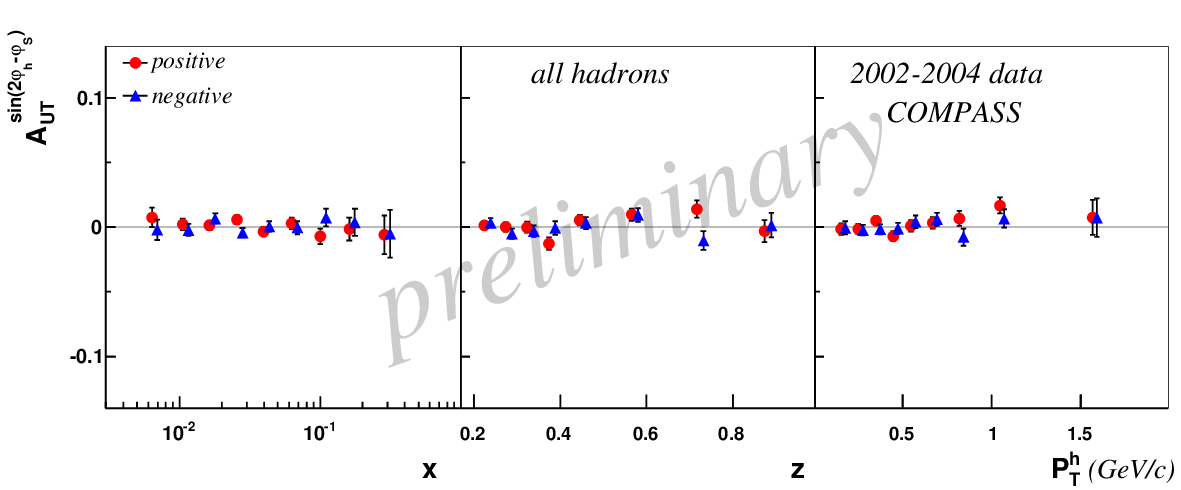} }
\caption{$A_{UT}^{sin(2\phi_h-\phi_s)}$ asymmetry for positive (red
circles) and negative (blue triangles) hadrons vs. $\it x$, $z$ and
$P_{hT}$.}
\label{fig:3}       
\end{figure}

\begin{figure}[ht!]
\center \resizebox{0.85\textwidth}{!} {
\includegraphics{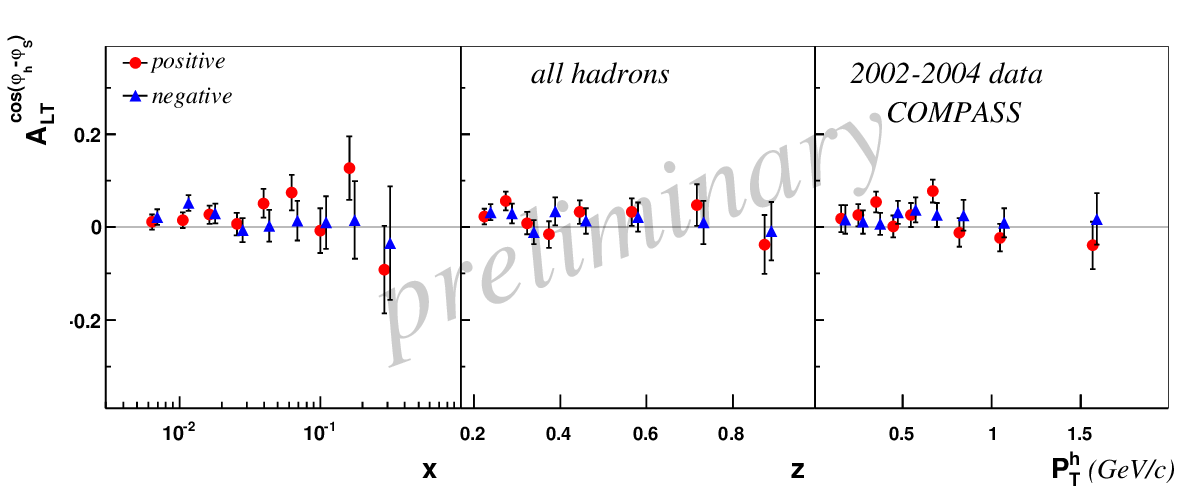} }
\caption{$A_{LT}^{cos(\phi_h-\phi_s)}$ asymmetry for positive (red
circles) and negative (blue triangles) hadrons vs. $\it x$, $z$ and
$P_{hT}$.}
\label{fig:4}       
\end{figure}

\begin{figure}[ht!]
\center \resizebox{0.85\textwidth}{!} {
\includegraphics{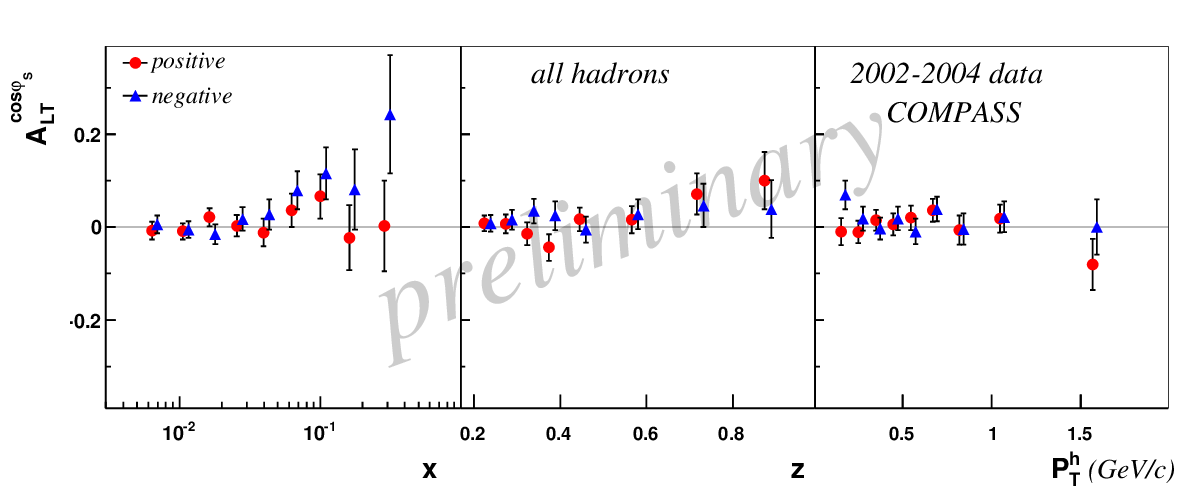} } \caption{$A_{LT}^{cos\phi_s}$
asymmetry for positive (red circles) and negative (blue triangles)
hadrons vs. $\it x$, $z$ and $P_{hT}$.}
\label{fig:5}       
\end{figure}

\begin{figure}[ht!]
\center \resizebox{0.85\textwidth}{!} {
\includegraphics{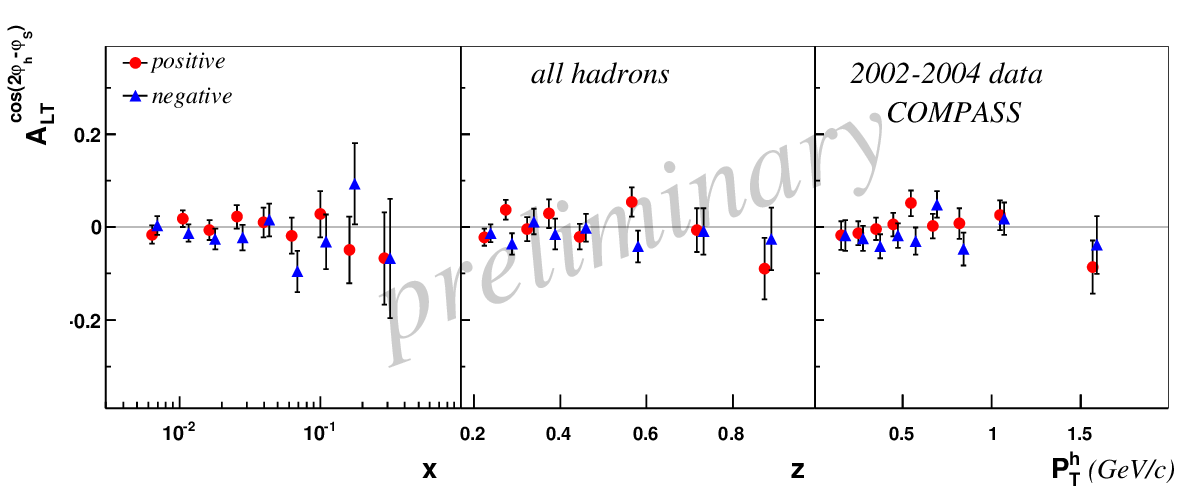} }
\caption{$A_{LT}^{cos(2\phi_h-\phi_s)}$ asymmetry for positive (red
circles) and negative (blue triangles) hadrons vs. $\it x$, $z$ and
$P_{hT}$.}
\label{fig:6}       
\end{figure}

We used for one measurement period (i.e. two subperiods with
opposite spin direction) the information of both target cells ($ \it
u,d$) and both sub-periods simultaneously by constructing the
estimator:
\begin{equation}\label{eq:dr_definition}
R(\Phi_j) = \frac {N^{+}_{u}(\Phi_j)N^{+}_{d}(\Phi_j)} {N^{-}_{u}(\Phi_j)N^{-}_{d}(\Phi_j)},\\
\end{equation}
It can be shown, that under a reasonable assumption on the ratio of
acceptances of the upstream and downstream cells to be constant
after the spin reversal $a^{+}_{u}(\Phi_j)/a^{-}_{d}(\Phi_j)$ $=$
$a^{-}_{u}(\Phi_j)/a^{+}_{d}(\Phi_j)$, the acceptance differences in
two cells cancel out, so finally one can obtain:

\begin{equation}\label{eq:dr_simple}
R(\Phi_j) = const (1+ 4W_j(\Phi_j))
\end{equation}
and the asymmetries can be extracted by fitting the $R(\Phi_j)$ with
the appropriate function. In Figs.~\ref{fig:1} - \ref{fig:6} we
present six target transverse spin dependent asymmetries extracted
for the first time from COMPASS 2002--2004 data collected on a
deuteron target. Asymmetries for positive and negative hadrons are plotted as a function of $x$, $z$ and $P_{hT}$.

\begin{figure}[ht!]
\center \resizebox{0.8\textwidth}{!} {
\includegraphics{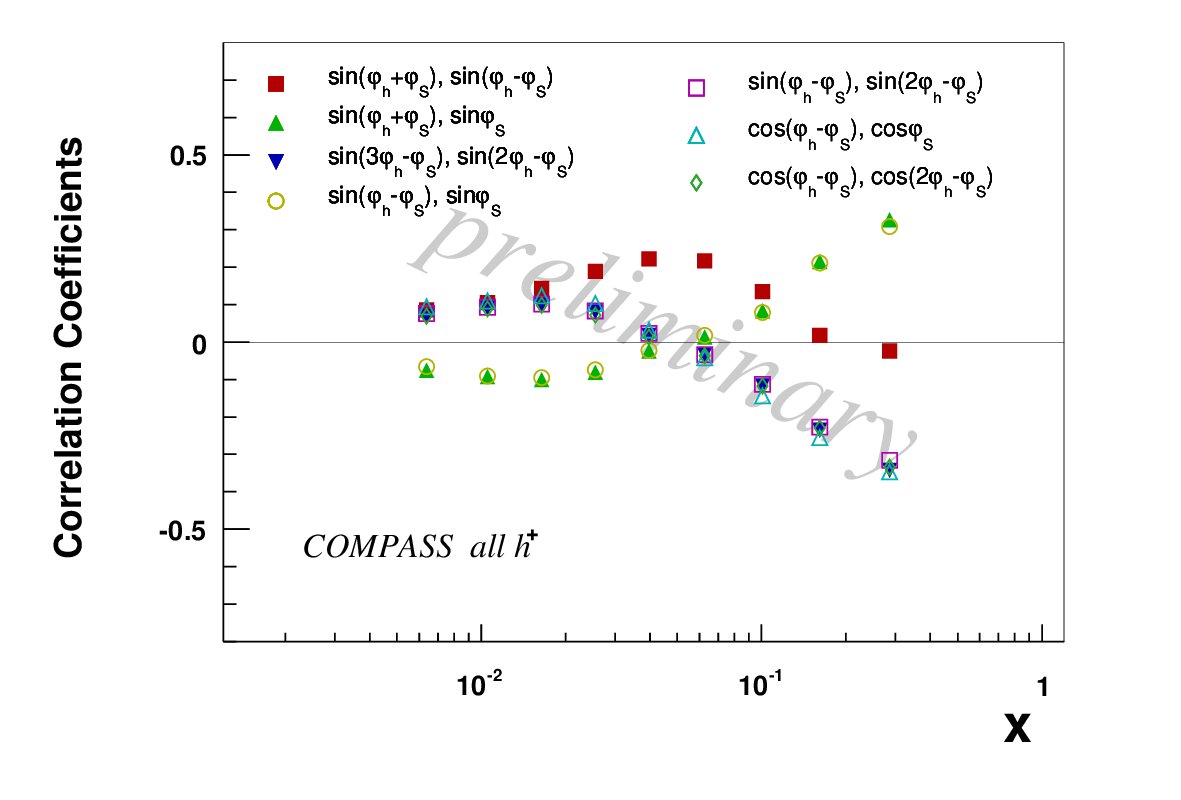} }
\caption{Correlation coefficients with the absolute value $>$ 0.1
vs. $\it x$.}
\label{fig:7}       
\end{figure}

Constructing the same estimator in two dimensional
$\varphi_h,\varphi_S$ space all the eight transverse spin
asymmetries can be extracted simultaneously by using the
two-dimensional fitting procedure. In addition this method reveal
possible correlations between the asymmetries. Corresponding checks
have shown only negligible or small correlations, as an example we
present in Fig.~\ref{fig:7} the correlation coefficients with
absolute values larger than 0.1 versus $x$, and it can be clearly
seen that even at maximum they remains smaller than 0.4. Various
systematic checks have been applied for the analysis. Finally the
estimated systematic errors are smaller than statistical ones.

All the six newly measured in COMPASS with deuteron target transverse spin asymmetries appear to be small. The smallness of azimuthal effects for deuteron target is interpreted by the different models predicting partial cancelation of u- and d- quarks contributions into the asymmetry.
Although the measured deuteron asymmetries are small this in no way affects the significance of the obtained result. As it was shown in recent "global" analysis by Anselmino et. al. ~\cite{Anselmino:2007fs} the d-quark distribution functions cannot be well defined without using COMPASS deuteron data.

\begin{figure}[ht!]
\center \resizebox{0.8\textwidth}{!}
{\includegraphics{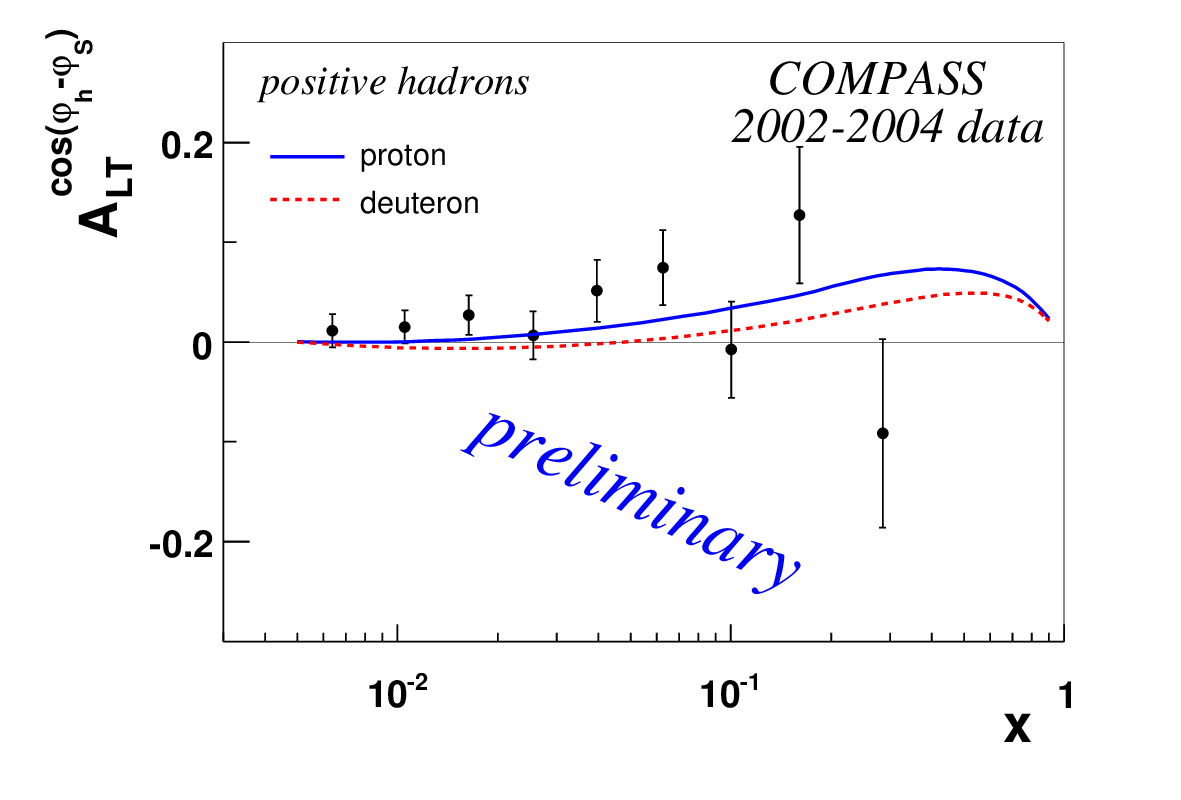} }
\caption{$A_{LT}^{cos(\phi_h-\phi_s)}$ asymmetry, positive hadrons
vs. $\it x$.}
\label{fig:8}       
\end{figure}

Our results for the $A_{LT}^{cos(\phi_h-\phi_s)}$
asymmetry have been compared with the predictions presented in ~\cite{Kotzinian:2006dw}. The authors performed calculations for the
$A_{LT}^{cos(\phi_h-\phi_s)}$ asymmetry by using some models for the
otherwise unknown $g_{1T}^q$ function and expressing it through the
well known integrated helicity distributions. In Fig.~\ref{fig:8} we
compare the curves plotting the calculated in
~\cite{Kotzinian:2006dw} $x$-dependence of the
$A_{LT}^{cos(\phi_h-\phi_s)}$ asymmetry in the COMPASS kinematical
region, with our experimental measurements.

The blue line
corresponds to the asymmetry calculated for the proton target and
the red dashed line is for the deuteron target. Experimental
observations do not contradict the predictions, and the theoretical
curve lies within experimental error bands. The asymmetry on proton target is
predicted to be roughly twice larger at high $x$ for COMPASS and even larger for HERMES and JLab kinematics. It would be very interesting to perform such measurements.

\section{Conclusions}\label{concl}
We have presented six new target transverse spin dependent
asymmetries extracted from COMPASS 2002--2004 data collected on a
deuteron target. The estimated systematic errors are smaller than
the statistical ones. All six newly measured asymmetries are
small, which is in agreement with the different models predictions. COMPASS has already started data taking with the proton target, and the transverse spin
effects, in this case, are expected to be more significant.
%
%
%

%
%
\end{document}